\documentclass[aps,prl,showpacs,preprint,superscriptaddress]{revtex4}
\usepackage{epsfig}
\begin{document}


\title{Giant Dipole Resonance Width in near-Sn Nuclei at Low Temperature and High Angular Momentum}

\author{Srijit Bhattacharya}
\affiliation{Darjeeling Government College, Darjeeling-734 101, India}
\author{S. Mukhopadhyay}
\author{Deepak Pandit}
\author{Surajit Pal}
\affiliation{Variable Energy Cyclotron Centre, 1/AF Bidhan Nagar, Kolkata-700 064, India}
\author{A. De}
\affiliation{Raniganj Girls' College, Raniganj-713 347, India}
\author{S. Bhattacharya}
\author{C. Bhattacharya}
\author{K. Banerjee}
\author{S. Kundu}
\author{T.K. Rana}
\author{A. Dey}
\author{G. Mukherjee}
\author{T. Ghosh}
\author{D. Gupta}
\author{S. R. Banerjee}
\email[e-mail:]{srb@veccal.ernet.in}
\affiliation{Variable Energy Cyclotron Centre, 1/AF Bidhan Nagar, Kolkata-700 064, India}


\date{\today}

\begin{abstract}

High energy $\gamma$-rays in coincidence with low 
energy yrast 
$\gamma$-rays have been measured from $^{113}$Sb, at 
excitation energies of 
109 and 122 MeV, formed by bombarding $^{20}$Ne on 
$^{93}$Nb at 
projectile energies of 145 and 160 MeV respectively 
to study the 
role of angular momentum (J) and temperature (T) over 
Giant Dipole Resonance (GDR) width ($\Gamma$). 
The maximum populated angular momenta for fusion were 
67$\hbar$ 
and 73$\hbar$ respectively for the above-mentioned beam
 energies. 
The high energy photons were detected using a Large Area
 Modular 
BaF$_2$ Detector Array (LAMBDA) along with a 24-element 
multiplicity 
filter. After pre-equilibrium corrections, the excitation 
energy E$^*$ was 
averaged over the decay steps of the compound nucleus (CN). 
The average 
values of temperature, angular momentum, CN mass etc. have 
been calculated 
by the statistical model code CASCADE. Using those average
 values,  
results show the systematic increase of GDR width with T
 which is 
consistent with Kusnezov parametrization and the Thermal
 Shape Fluctuation Model. 
The rise of GDR width with temperature also supports the 
assumptions of 
adiabatic coupling in the Thermal Shape Fluctuation Model. 
But the 
GDR widths and corresponding reduced plots with J are 
not consistent 
with the theoretical model at high spins. 
\end{abstract}

\pacs{24.30.Cz, 24.60.Dr, 25.70.Gh, 27.60.+j}


\maketitle

\section{introduction}
The study of nuclear structure and dynamics under 
extreme conditions of 
internal energy and angular momentum is important in understanding
the diverse properties of atomic nuclei. 
The measurements of high energy gamma rays emitted 
when isovector Giant Dipole Resonances (GDR) 
in highly excited nuclei are damped can provide information
 on the various nuclear
properties at finite temperature (T) and 
angular 
momentum (J)~\cite{Gaar,Snov}. Although the excited 
state GDR in 
heavy ion fusion reactions was observed in the early
 eighties, 
the study of this resonance still continues to be a very 
interesting 
and useful tool in the field of nuclear structure and 
dynamics~\cite{Thon}. 
The systematics of GDR width ($\Gamma$) as a function 
of T and J still 
remains a hotly debated and puzzling topic. The central 
issue is to understand 
the role of different damping mechanisms viz. collisional 
damping~\cite{Bar} 
and adiabatic thermal shape fluctuation~\cite{Ormond,Orm} 
with their 
dependence on T and rotational frequency of the nucleus.

Till now most of the measurements of GDR cross-section 
built on excited states 
have been made with Sn and near-Sn nuclei formed by 
heavy ion fusion reactions. 
Previous measurements~\cite{Bor,Gaar,Brac,Hof,Pier} 
suggest the continuous 
growth of GDR width with excitation energy ($E^*$) 
up to 
120-130 MeV (T $\leq$ 2 MeV) and attribute the same 
to rapid 
increase of spin-induced deformations and thermal shape 
fluctuations. As 
per the experimental observations, beyond the bombarding 
energy at 
which angular momentum saturates, the increase in GDR width
 is very small. This 
saturation of width is interpreted as the evidence for the
 onset of 
maximum angular momentum the nucleus can sustain without 
fissioning. Theoretically, 
the Thermal Shape Fluctuation Model (TSFM), in general, 
can predict the 
trend of the experimental data for $E^* \leq$ 120-130 MeV
 but after that 
it fails to show any saturation of GDR width with 
increasing E$^*$. On the contrary, 
the effect of temperature on GDR width, as is revealed 
from the work 
of Kelly et al.~\cite{Kelly} is quite inconsistent with 
the saturation 
previously observed. Kelly et al. emphasized that at 
higher bombarding 
energies the excitation energy and temperature should 
be correctly 
redefined. Their work results in a very interesting 
observation that 
if nuclear temperature is estimated using average E$^*$ 
after proper 
pre-equilibrium corrections, previous results could 
also indicate the 
increase of GDR width up to T $\leq$ 3.5 MeV
(unlike up to 
2 MeV as previoulsy observed). The general trend of
 those experimental 
data agrees well with the predictions of the TSFM. In spite
 of that, 
some recent observations, in the region T $\leq$ 2 MeV, 
show that 
the experimental findings of GDR width are smaller 
than the 
predictions of TSFM in at least 4 different mass 
regions (Cu, Sn, Pb, Au)~\cite{Thon,Kus,Bau,Heck}.

Very recently, the study of angular momentum dependence 
of 
$\Gamma$ keeping $E^*$ unchanged has become more 
controversial. Previous
 experimental findings agree with the fact that 
$\Gamma$ remains 
constant up to J $\sim$ 30-40$\hbar$ which is in 
conformity with 
Kusnezov parametrization of TSFM. However, at higher 
J values a 
few recent results seem to violate this parametrization. 
In the region
 T $\leq$ 2 MeV and with higher angular momenta, the 
TSFM can explain 
the variation of $\Gamma$ in the case of $^{106}$Sn and 
$^{176}$W~\cite{Bracco,Mati} 
but fails for $^{86}$Mo to do so~\cite{Rathi}. The recent 
investigation 
by Chakrabarty~\cite{DRC} emphasizes that the average 
values of 
temperature, angular momentum and mass must be smaller 
than those of 
the initial compound nucleus considered. Interestingly, 
if the average values of those parameters are taken into 
account, 
Kusnezov parametrization can successfully explain the 
experimental 
data of $^{86}$Mo but it fails in the case of Sn~\cite{DRC}. 
Thus 
though the model of thermal shape fluctuation describes 
rather well, 
on the average, many experimental results, it fails to 
reproduce the 
data corresponding to the lowest temperatures and  
highest spins.

Under these circumstances, it comes out to be that plenty of 
experimental 
data are needed in the region of T $\leq$ 2 MeV with higher 
angular momenta 
to understand the limits of the TSFM and Kusnezov
 parametrization.

 The present experiment revisits excited state GDR 
in $^{113}$Sb - a near 
Sn nuclide, to provide new results at higher spins 
(in the region of 40-60 $\hbar$) 
in order to test the simple parametrization given by 
Kusnezov,  which is 
based on, mainly, the data at low and medium spins. 
The values of E$^*$, J 
and mass (A) have been estimated by averaging over the 
decay steps of the 
compound nucleus.  In this work an effort is made to 
provide the temperature 
of the emitting nucleus applying all the necessary 
corrections including that 
of the pre-equilibrium emission.  The pre-equilibrium 
corrections have been 
done using the parametrization of Kelly et al.~\cite{Kelly}.

\section{Experimental details}
\begin{figure}
\includegraphics[height=6 cm, width=8 cm]{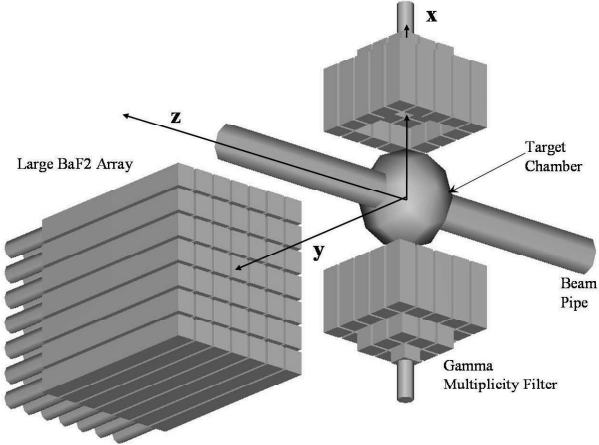}%
\caption{\label{fig1}Schematic view of experimental set-up for the 
LAMBDA (Large BaF$_2$ array) spectrometer in a 7x7 matrix arrangement 
along with the low energy $\gamma$-ray multiplicity filter.}
\end{figure}

 The experiment was performed at the Variable 
Energy Cyclotron Centre (VECC), 
Kolkata. A 1 mg/cm$^2$ thick target of 99.9\% 
pure $^{93}$Nb was bombarded 
with a beam of $^{20}$Ne produced by the K130 
Cyclotron of VECC. Two different
 beam energies of 145 MeV and 160 MeV were 
employed forming the compound 
nucleus $^{113}$Sb at the excitation energies 
of 109 MeV and 122 MeV respectively. 
The maximum populated angular momenta for fusion were 
67$\hbar$ and 73$\hbar$ respectively 
at the two bombarding energies. The inclusive 
high energy $\gamma$-rays were 
detected with a part of the BaF$_2$ detector 
array LAMBDA~\cite{Sri1,NIM}. The 
array comprised of 49 detectors arranged in 
a 7$\times$7 square matrix 
configuration, each detector having a length 
of 35 cm and a square face 
of 3.5$\times$3.5 cm$^2$ area. The detector 
array was positioned at a distance 
of 50 cm from the target and at an angle of 
$55^{\circ}$ w.r.t. the beam axis. The 
array subtended a solid angle of 
0.227 sr (1.8\% of 4$\pi$). Lead sheets 
of 3 mm 
thickness were placed in front and sides 
of the array to cut down the low energy 
$\gamma$-rays and X-rays. The beam dump 
was heavily shielded with borated paraffin 
and lead bricks to decrease the neutron and 
$\gamma$-ray background. A 24-element 
multiplicity filter detector array was used 
along with the LAMBDA spectrometer to 
 measure the multiplicity of low energy 
$\gamma$-rays in coincidence with the high 
energy $\gamma$-rays. The multiplicity detector 
assembly also consists of BaF$_2$ 
detectors, each 
3.5 $\times$ 3.5 $\times$ 5.0 cm$^3$ in 
dimension, packed in two 
groups of 12 each and placed on the 
two sides of the target chamber at a distance 
of 10 cm from the target. The complete 
detection system is shown schematically in 
the Fig. \ref{fig1}. 

The response of the LAMBDA spectrometer~\cite{NIM} 
was generated using the Monte 
Carlo Code GEANT 3.21~\cite{CERN} 
incorporating realistic geometry of 
the array, the energy resolution of the 
detectors, experimental 
conditions of shielding, discriminator 
thresholds etc. The energy 
calibration of the individual detectors 
was done using the low energy 
$\gamma$-ray sources viz. 
$^{22}$Na (0.511 MeV, 1.274 MeV, and the 
sum peak 1.785 MeV), $^{60}$Co (Sum Peak 2.505 MeV), 
$^{241}$Am-$^9$Be (4.43 MeV)
 and also by minimum ionising peak (23.1 MeV) of 
cosmic muons. The energy
 response of the detectors was found to be linear 
up to 23.1 MeV. The time 
resolution of individual detectors was 
960 ps. The response of 
the $\gamma$-ray multiplicity detector array was 
also generated using 
GEANT 3.21 simulation code~\cite{Sri}. 
A dedicated electronics setup (consisting of 
multi-inputs CAMAC and NIM 
modules and a VME based data acquisition system 
capable of handling 
$\sim$ 4K events/second without appreciable 
dead-time loss) was used to 
register the energy and time from each detector 
in an event by event mode. An 
event was treated as a valid event only when the 
deposited energy in any 
detector crosses a high threshold (T$_h$) of 
4 MeV. 
 The details of high energy gamma spectrometer 
LAMBDA, its response, electronics 
setup and event reconstruction method have 
already been described in 
S.Mukhopadhyay et al.~\cite{NIM}. 

\begin{figure}
\includegraphics[height=6 cm, width=8 cm]{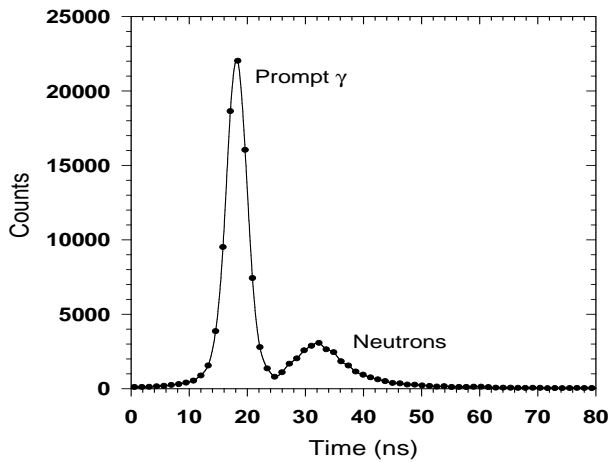}%
\caption{\label{fig2} The experimental time spectrum 
obtained from a 
single detector showing a clear separation between 
the neutrons and prompt $\gamma$-rays.}
\end{figure}

Time of Flight (TOF) technique was used to 
eliminate neutrons. For TOF 
measurements, the time reference was taken from
 the multiplicity 
detector assembly. A clear separation between the 
neutrons and 
prompt $\gamma$-rays was seen for all the detector
 elements in the experimental time spectra (Fig. \ref{fig2}). 
Pulse Shape Discrimination (PSD) method by 
long (2 $\mu$s) -- short (50 ns) 
integration technique was adopted to reject pile-up 
events. Since 
the detector array is highly segmented, pile-up 
events were very small.  
In triggered data acquisition mode the probability 
of cosmic events 
is small (rejection ratio better than 1:3300) and 
also those events 
can be rejected effectively from the hit pattern 
utilizing the 
square detector geometry and high segmentation of 
the array.

\section{Experimental Data Analysis}
The high energy $\gamma$-ray spectra were generated 
from the 
event-by-event data during offline analysis. For 
reconstructing 
the events, a nearest neighbor (cluster) summing 
technique was adopted. In 
this technique, first, the detector with highest 
energy deposited above a 
high threshold ($\geq$ 4 MeV) within the array was 
identified and named 
as primary detector. For obtaining the full energy 
information of the 
incident photons, it is important to confine the 
secondary electromagnetic 
shower within the array volume as far as possible. 
Therefore, in the above-mentioned 
event reconstruction technique a checking was done 
to find whether the primary 
detector was surrounded by all its neighbors i.e., 
all the 
8 elements (irrespective of any hits in them). The 
event was treated as 
valid, if this condition was 
satisfied. Otherwise, due to the 
possibility of losing a part of the electromagnetic 
shower, the event 
was rejected. In the case of a valid event, for the 
final adding back, only 
those detectors were considered among the 8 nearest 
neighbors, having 
an energy deposit $>$ 250 keV. The same scheme had been 
adopted while 
simulating the response of the array using GEANT. During 
the adding back, 
the hit events in the individual elements in the cluster 
were validated by the prompt 
gamma cuts in TOF spectra and long-short PSD selections.  
Next by gating on different coincidence folds of low energy 
$\gamma$-multiplicities 
in the multiplicity array the high energy $\gamma$-ray 
spectra were generated for 
each beam energy. The contributions due to the chance 
coincidence events within the
 prompt $\gamma$ window in the TOF spectrum were also 
subtracted. Finally the 
spectra were Doppler corrected. 

The conversion between the measured coincidence fold 
$F_\gamma$ (the number of 
measured coincident $\gamma$-rays of low energy in each event) 
to the 
multiplicity $M_\gamma$ (the number of $\gamma$-rays emitted 
in the reaction) 
was established using the response matrix $S(F_\gamma, M_\gamma)$. 
This 
response matrix was generated by making use of the GEANT, 
where, realistic 
multiplicity detector setup was considered and low energy 
$\gamma$-rays were 
thrown isotropically with incident multiplicity distribution 
P(M). It has 
been assumed that the multiplicity distribution following 
fusion reaction 
is given by~\cite{Mati},

 \centerline{P(M) = M$_\gamma/[1+exp\{(M_\gamma-M_0)/\delta M\}]$}

\begin{figure}
\includegraphics[height=6 cm, width=8 cm]{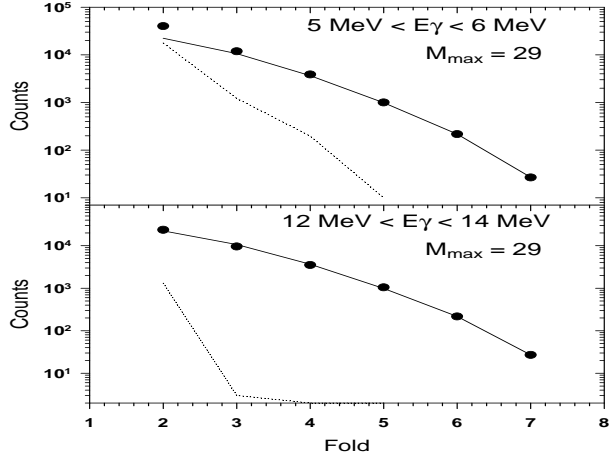}%
\caption{\label{fig3} Experimental fold spectrum fitted 
with GEANT 
simulation (solid line) for  the high energy $\gamma$ 
windows 
5 MeV $ < E_\gamma < $ 6 MeV (top panel) and for 
12 MeV $ < E_\gamma < $ 14 MeV 
(bottom panel) for beam energy = 160 MeV. The dotted line 
is the difference 
between the measured and predicted fold distributions.}
\end{figure}

The maximum of multiplicity M$_0$ and diffuseness $\delta$M 
were obtained 
by fitting the equation, 

\begin{displaymath}
\sum{S(F_{\gamma}, M_{\gamma})P(M_{\gamma})} = f_{exp}(F_{\gamma})
\end{displaymath}

where f$_{exp}(E_\gamma)$ is the measured multiplicity 
spectrum in the 
region 2 $\leq$ F$_\gamma \leq $7. Typical experimental 
fold distributions 
are shown in the Fig. \ref{fig3}. 

\begin{figure}
\vspace{ 1.5 cm}
\includegraphics[height=6 cm, width=8 cm]{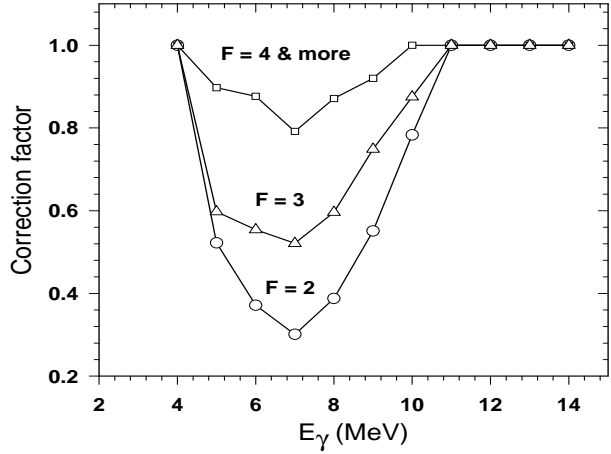}%
\caption{\label{corr} The correction factors are plotted 
against $\gamma$-ray energies for fold 2 (circles), 
3 (triangles) and $\geq$ 4 (squares) in the case of E$_{beam}$ = 160 MeV. 
The decrease of correction factors (see text for details) with
 the increase of fold is evident.}
\end{figure}

\begin{figure}
\vspace{ 1.5 cm}
\includegraphics[height=6 cm, width=8 cm]{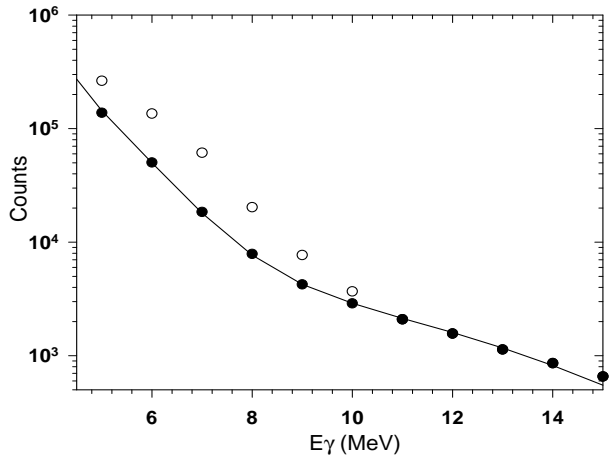}%
\caption{\label{figcor} High energy $\gamma$-ray spectra 
(expanded up to 15 MeV)
 corresponding to fold 2 at 160 MeV beam energy before 
(shown by open circles) 
and after corrections (shown by filled circles). Corresponding 
statistical 
model fit is shown by solid lines. The error bars of the 
respective points 
are less than the size of the symbols.}
\end{figure}

The theoretical fold distributions were found to be matching 
well with the 
experimental fold distributions for 145 and 160 MeV for 
$E_{\gamma}$ $\geq$ 12 MeV.  
For 4 MeV $\leq$  E$_\gamma$  $<$ 12 MeV, another low 
multiplicity component 
had to be added. This actually is the difference between 
experimentally 
measured and theoretically predicted fold distributions 
and is shown by the 
dotted line in the Fig. \ref{fig3} . The intensity of the 
low multiplicity 
component peaks at around 7 MeV and falls on either side 
becoming negligible 
beyond 12 MeV. Its contribution was estimated for each 
fold from the fits 
of the fold distributions. The correction factors for the 
high energy 
$\gamma$-ray spectra were generated from the percentage 
contribution of the 
above-mentioned low multiplicity component. The correction 
factors are 
plotted in the Fig. \ref{corr} against the $\gamma$-ray 
energies corresponding 
to different folds for 160 MeV incident energy. The corrections 
decreased as 
a function of fold and became negligibly small for fold F $\geq$ 4.  The 
high energy $\gamma$-ray spectra corresponding to selected 
folds were 
multiplied (in the region below 12 MeV) by the energy dependent 
correction factors 
and corrected accordingly. The Fig. \ref{figcor} 
shows a typical 
high energy $\gamma$-ray spectrum (for fold = 2 at 160 MeV projectile energy) 
thus corrected (filled circles) along with the raw 
uncorrected one (open circles). 
The measured enhanced yield at low folds could be due to 
non-fusion events. 
Since the reaction studied has an asymmetry between N/Z of 
the target and 
projectile, there might be a possibility of pre-equilibrium 
$\gamma$-ray
emissions as those are related to the dynamic dipole formation 
in the 
fusion entrance channel.  However, since the recoiling nucleus 
was not 
directly measured it is not possible to disentangle the different 
effects.

From the multiplicity distributions for different folds, the 
angular 
momentum (J) distributions of the compound nucleus have been 
extracted 
and the average value of J (J$_{CN}$) for the compound nucleus 
was 
calculated. The conversion from multiplicity to angular 
momentum was 
done assuming J = 2M+k, k = 4 takes into account the angular 
momentum 
removed by nonstatistical gamma rays, particle emission and 
gamma rays 
below trigger thresholds. The fitted values of parameters 
$\delta$M, $M_{max}$, $J_{max}$ with corresponding beam 
energies 
are shown in Table \ref{tab:mult}. The average angular 
momenta 
and the corresponding widths for different folds are shown 
in Table \ref{tab:delj}.

\begingroup
\begin{table}
\caption{\label{tab:mult} Table showing the fitted values of 
parameters of multiplicity and angular momentum distributions.}
\begin{ruledtabular}
\begin{tabular}{|c|c|c|c|}
  E$_{proj}$ & M$_{max}$ & $\delta$M & J$_{max}$ \\ 
  (MeV) &  &  & ($\hbar$) \\ 
\hline
  145 & 28 & 3 & 60\\
  160 & 29 & 3 & 62\\
\end{tabular}
\end{ruledtabular}
\end{table}
\endgroup

\begingroup
\begin{table}
\caption{\label{tab:delj} Table showing the angular 
momenta and corresponding widths for different folds 
at two beam energies.}
\begin{ruledtabular}
\begin{tabular}{|c|c|c|c|}
  E$_{proj}$ & fold & J$_{CN}$ & FWHM in J \\ 
  (MeV) & ($\hbar$)& ($\hbar$)& ($\hbar$) \\ 
\hline
  145 & 2 & 49 & 24\\
  145 & 3 & 53 & 22\\
  145 & 4 & 57 & 18\\
  145 & $\geq$5 & 59 & 16\\
  160 & 2 & 50 & 24\\
  160 & 3 & 54 & 20\\
  160 & $\geq$4 & 59 & 18\\
\end{tabular}
\end{ruledtabular}
\end{table}
\endgroup

\subsection{Statistical Model Analysis}
The high energy $\gamma$-ray data collected in the experiment were 
sorted into 
4 different classes corresponding to folds 2, 3, 4 and $\geq$ 5. The
 measured high energy $\gamma$ spectra associated with different 
folds were fitted with a modified version of statistical model 
code CASCADE~\cite{Pul} along with a bremsstrahlung component. 
There
 is enough experimental evidence that at projectile energy above 
6 MeV/u, 
contribution of pre-equilibrium particle emission becomes 
important and 
should be included in CASCADE. For calculation of proper 
E$^*$, 
pre-equilibrium estimates were done on the basis of the 
empirical 
formula~\cite{Kelly}
 
\centerline{$\Delta E_{x}(MeV) = 8.7[(E_{proj}-V_c)/A_{proj}]-33$}

 where $V_c$ is the Coulomb barrier. This parametrization 
based on the
 demonstrated scaling with $(E_{proj}-V_c)/A_{proj}$ 
(insensitive to 
the target-projectile combinations~\cite{Tri}) have been 
used in 
this work to estimate the energy lost in pre-equilibrium 
emission 
and to correct the excitation energy ($V_c$ = 57.0 MeV at 
r = r$_c$ = 10.35 fm). 
Corresponding pre-equilibrium energy loss is 5.28 MeV at 
$E_{lab}$ = 145 MeV (5\% of initial excitation energy) and 
11.8 MeV at $E_{lab}$ = 160 MeV (10\% of initial excitation 
energy). The corrected excitation energy was used within 
CASCADE, in which Reisdorf 
level density prescription~\cite{Reis} 
had been used. The asymptotic level density parameter~\cite{dios} 
was taken as $\tilde{a}$ = A/8.0 MeV$^{-1}$. In the statistical 
model calculation, a single lorentzian GDR strength function was 
assumed, 
having centroid energy (E$_{GDR}$), strength (S), width 
($\Gamma$) 
as parameters .  In the CASCADE calculation, the moment of 
inertia I of 
the compound nucleus was taken as,  
${\cal I} = {\cal I}_0(1+\delta_1J^2+\delta_2J^4)$ 
where ${\cal I}_0$ is the spherical moment of inertia. 
The parameters $r_{eff}$, $\delta_1$ and $\delta_2$ were 
kept 
at default values of 1.22 fm, $0.4699 \times 10^{-5}$ and 
$0.9326 \times 10^{-8}$ respectively within the CASCADE. The 
spin 
distributions for different folds of the compound nucleus 
deduced 
from the experimental multiplicity distribution were used as 
inputs
 in CASCADE. The critical angular momenta $l_{cr}$, 
fusion 
cross sections and the corresponding projectile 
energies 
are shown in Table \ref{tab:lcr}.
\begin{table}
\caption{\label{tab:lcr}Table showing critical angular 
momenta, 
fusion cross-sections for each beam energy calculated by 
CASCADE.}
\begin{ruledtabular}
\begin{tabular}{llcrr}
  & E$_{beam}$ & J$_{cr}$ & $\sigma _{fus}$ \\
  & (MeV) & ($\hbar$) & (mb) \\ \hline
  & 145 & 67 & 1390 \\
  & 160 & 73 & 1450 \\
\end{tabular}
\end{ruledtabular}
\end{table}

\subsection{Bremsstrahlung Contribution}

\begin{figure}
\vspace{1.5 cm}
\includegraphics[height=8 cm, width=12 cm]{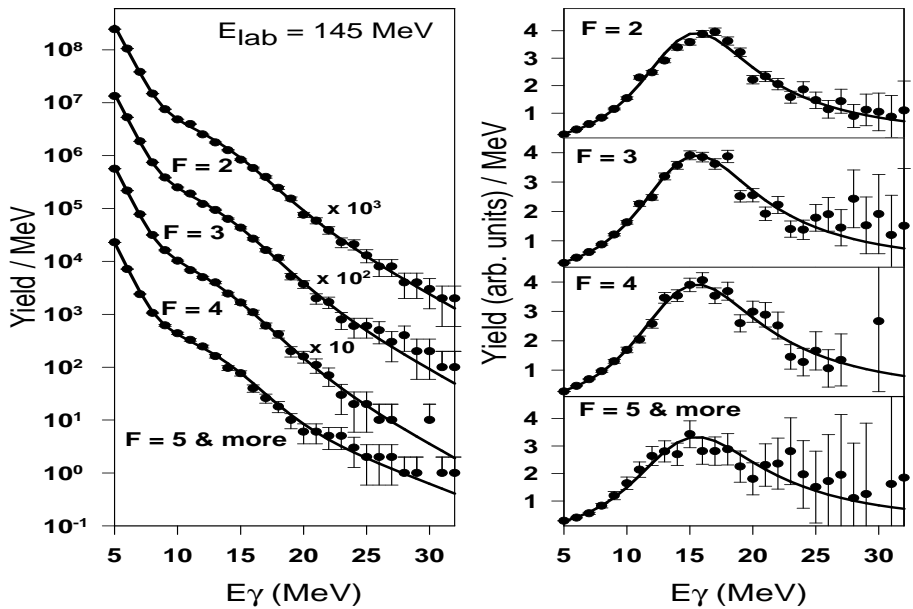}%
\caption{\label{fignb145} Left: high energy $\gamma$-ray spectra 
for different folds for beam energy 145 MeV are plotted, 
right: the linearized GDR spectra for different folds are 
plotted with $\gamma$-ray energies.}
\end{figure}

\begin{figure}
\includegraphics[height=8 cm, width=12 cm]{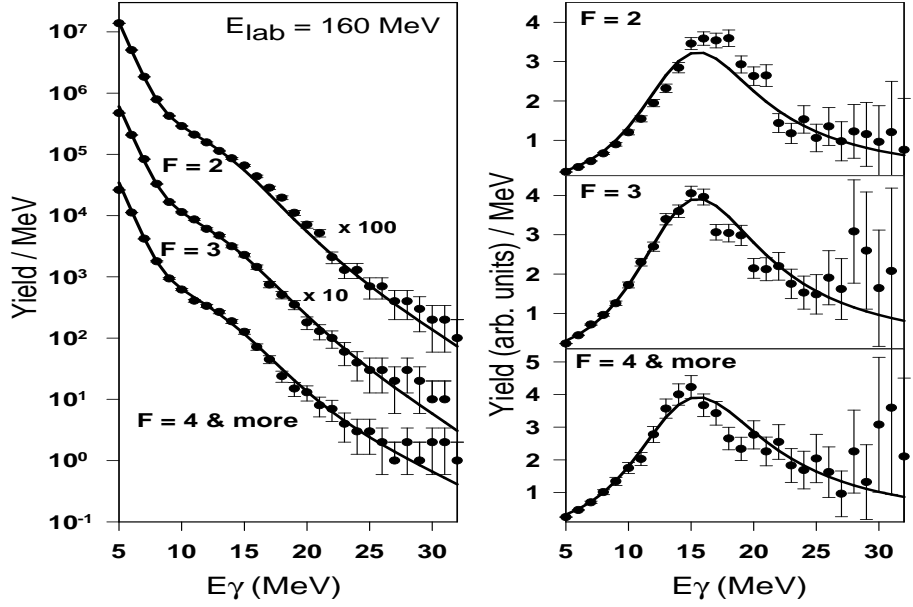}%
\caption{\label{fignb160} Left: high energy 
$\gamma$-ray spectra for different folds at 
beam energy 160 MeV, right: the linearized GDR spectra 
for different folds plotted against $\gamma$-ray energies.}
\end{figure}

The non-statistical 
contributions to the experimental 
$\gamma$-ray 
spectra arising from bremsstrahlung processes were assumed 
to have 
an energy dependence of $exp(-E_{\gamma}/E_0)$, where the 
slope 
parameter E$_0$ was chosen according to bremsstrahlung 
systematics~\cite{Nife,Ram}. 
The contribution was normalized to the experimental spectra 
at 
25-30 MeV and was added to the calculated $\gamma$-ray 
spectra 
from CASCADE after folding with detector response. 
The measured $\gamma$-ray spectra were fitted within 
the 
region E$_\gamma$ = 8-25 MeV with CASCADE using a 
$\chi^{2}$ minimisation routine. The experimental 
spectra 
fitted with CASCADE plus bremsstrahlung for different 
folds 
are shown in figures \ref{fignb145} and \ref{fignb160} 
for E$_{beam}$ = 145 and 160 MeV respectively . 
The 
linearized GDR plots 
were extracted by the transformation 
$f(E_{\gamma})*Y_{\gamma}^{exp} / Y_{\gamma}^{cal}$ 
for E$_{beam}$ = 145 and 160 MeV and shown in 
figures \ref{fignb145} and \ref{fignb160} respectively, 
where, $Y_{\gamma}^{exp}$ is the experimental spectrum 
and $Y_{\gamma}^{cal}$ is the prediction from CASCADE 
folded with the detector response. f(E$_\gamma$) is 
the GDR strength function and is given by,

\vspace{0.2cm}
\begin{displaymath}
f(E_{\gamma})=\frac{E_\gamma \Gamma_{GDR}}
{[(E_{\gamma}^2-E_{GDR}^2)^2+E^2_{\gamma} \Gamma_{GDR}^2]}
\end{displaymath}

 The values of GDR parameters are shown in Table \ref{tab:wid}. 

\begingroup
\squeezetable
\begin{table}
\caption{\label{tab:wid}Table showing GDR and bremsstrahlung 
parameters for different beam energies calculated by CASCADE 
as explained in the text.}
\begin{ruledtabular}
\begin{tabular}{|c|c|c|c|c|c|}
  E$_{beam}$ & Fold     & Strength    & E$_{GDR}$ &   $\Gamma_{GDR}$  & E$_0$  \\ 
  (MeV)      &   (F)    & (S$_{GDR}$) &   (MeV)   &    (MeV)       & (MeV)  \\ \hline
  145 & 2 & 1.01 $\pm$ 0.01 & 15.55 $\pm$ 0.10 & 11.5 $\pm$ 0.25 & 3.0\\
  145 & 3 & 0.98 $\pm$ 0.02 & 15.55 $\pm$ 0.10 & 11.8 $\pm$ 0.25 & 3.0\\
  145 & 4 & 0.99 $\pm$ 0.01 & 15.55 $\pm$ 0.12 & 12.4 $\pm$ 0.25 & 3.0\\
  145 & $\geq$5 & 0.98 $\pm$ 0.02& 15.55 $\pm$ 0.12 & 12.8 $\pm$ 0.25 & 3.0\\
  160 & 2 & 0.98 $\pm$ 0.02 & 15.55 $\pm$ 0.12 & 11.9 $\pm$ 0.25 & 3.5\\
  160 & 3 & 0.98 $\pm$ 0.02 & 15.55 $\pm$ 0.12 & 12.5 $\pm$ 0.25 & 3.5\\
  160 & $\geq$4 & 0.98 $\pm$ 0.02 & 15.55 $\pm$ 0.12 & 13.0 $\pm$ 0.25 & 3.5\\
\end{tabular}
\end{ruledtabular}
\end{table}
\endgroup

\subsection{Temperature Estimation}
At high excitation energy the compound nucleus decays 
through a large number of decaying steps, and hence the
 mass (A), charge (Z), excitation energy (E$^*$) and 
angular momentum (J) of the compound nucleus should be 
averaged over all the decay steps. The average values of 
E$^*$, J, A, Z should be different and less than those 
of the initial compound nucleus.  
While estimating the average temperature, in 
accordance 
with the prescription adopted by Wieland et 
al.~\cite{Wieland}, 
a lower limit in excitation energy during the CN 
decay process 
was employed in the statistical model 
calculation. This cut 
in E$^*$ affects the low energy part of the high energy 
$\gamma$-spectra, 
without affecting the region of our interest 
$E_\gamma$ = 12-25 MeV. 
The average values of mass, atomic number, pre-equilibrium 
corrected 
excitation energy and angular momentum were calculated 
using the 
above-mentioned E$^*$ limit.  Within this E$^*$ limit, 
the estimated 
average values correspond to approximately 50\% of the 
total 
high energy $\gamma$-ray yield ($E_\gamma$ = 12-25 MeV) 
in CN 
decay chain. The average temperature was estimated from 
$\overline{E}^*$ by using the relation,

$\overline{T} = [(\overline{E}^*-\overline
{E}_{rot}-E_{GDR}-\Delta_p)/a(\overline{E}^*)]^{1/2}$
where $\overline{E}^*$ is the average of the 
excitation energy 
after pre-equilibrium subtraction weighed over the 
daughter 
nuclei for the $\gamma$-emission in the GDR region 
from $E_{\gamma}$ = 12-25 MeV:

$\overline{E}^*= \sum_{i}(E^*_i \omega _i)/\sum_{i}\omega_i$

$E_i^*$ is the excitation energy of i$^{th}$ nuclei 
in the 
decay steps and $\omega_i$ is the yield in the 
region 
$E_{\gamma}$ = 12-25 MeV. E$_{rot}$ is the energy 
bound 
in the rotation computed at the average J calculated 
within 
the CASCADE corresponding to analysed fold~\cite{Kim,ger}. 
E$_{GDR}$ is the centroid energy of the GDR, given 
by 
15.55 MeV and $\Delta _p $ is the pairing energy. 
The 
details of the parameters for different beam energies 
calculated by CASCADE for this experiment are shown 
in Table \ref{tab:kus}.

\begingroup
\squeezetable
\begin{table}
\caption{\label{tab:kus}Table showing parameters for 
different beam energies calculated by CASCADE for this experiment.}
\begin{ruledtabular}
\begin{tabular}{|c|c|c|c|}
\vline
  E$_{beam}$ & J$_{CN}$ & J$_{mean}$ & T$_{mean}$ \\ 
  (MeV)      &  ($\hbar$) & ($\hbar$) &   (MeV)  \\ \hline

  145 & 49.0 & 41.0 & 1.94$^{+0.06}_{-0.1}$ \\
  145 & 53.2 & 48.0 & 1.87$^{+0.06}_{-0.1}$ \\
  145 & 56.7 & 50.0 & 1.81$^{+0.03}_{-0.1}$ \\
  145 & 59.5 & 54.0 & 1.72$^{+0.07}_{-0.03}$ \\
  160 & 50.0 & 44.0 & 1.98$^{+0.14}_{-0.05}$ \\
  160 & 54.3 & 47.0 & 1.90$^{+0.13}_{-0.04}$ \\
  160 & 58.7 & 53.0 & 1.86$^{+0.09}_{-0.14}$ \\
\end{tabular}
\end{ruledtabular}
\end{table}
\endgroup

\section{Experimental Results}

\begin{figure}
\vspace{2 cm}
\includegraphics[height=7 cm, width=8 cm]{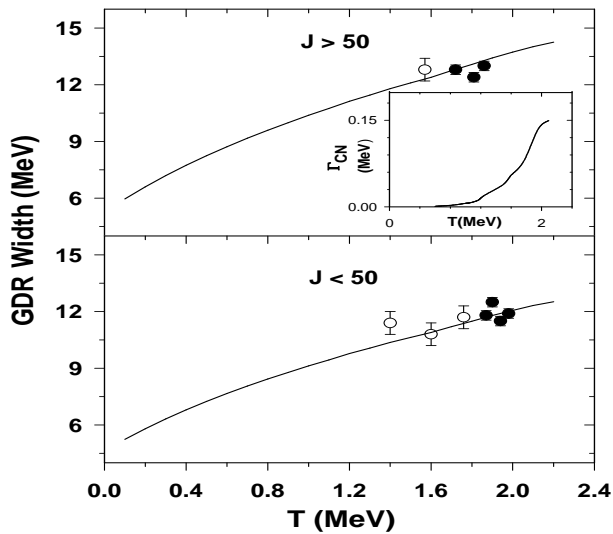}%
\caption{\label{figt} top:  GDR widths are plotted against T 
and compared with Kusnezov calculation for J $>$ 50$\hbar$,  
bottom:  Same plot for J $<$ 50$\hbar$. The filled circles 
denote experimental data from this work and the points with the 
open circles are from Bracco et al~\cite{Bracco}. 
Same scheme for averaging the T and J
has been adopted in both the cases (see text for details). inset: compound 
nuclear particle decay width plotted against T.}
\end{figure}

The GDR widths $\Gamma$(T, J, A) measured in this work are 
plotted against average
values of T. These are shown as filled circles
in the top panel of the Fig. \ref{figt} 
for angular momenta
greater than or equal to 50$\hbar$ and in the bottom 
panel for 
angular momenta less than
50$\hbar$ . The TSFM predictions using Kusnezov 
parametrization for
J $>$ 50$\hbar$ and J $<$ 50$\hbar$ are also represented 
by solid 
lines in the top and 
the bottom panels. The average mass in this case is 
111. 
TSFM calculation (indicated by solid line)
includes the average compound nucleus particle decay 
width 
calculated with an asymptotic
level density parameter $\tilde{a}$ = A/8.0 MeV$^{-1}$ . 
The graph in the inset of the
Fig. \ref{figt} shows the average compound nucleus 
particle 
decay width plotted against
temperature. In this E$^*$ range, the $\Gamma_{CN}$ 
has 
small magnitude and grows with
increase in T. The inclusion of $\Gamma_{CN}$ improves 
the 
fitting in Fig \ref{figt}
marginally.

\begin{figure}[t]
\includegraphics[height=8 cm, width=10 cm]{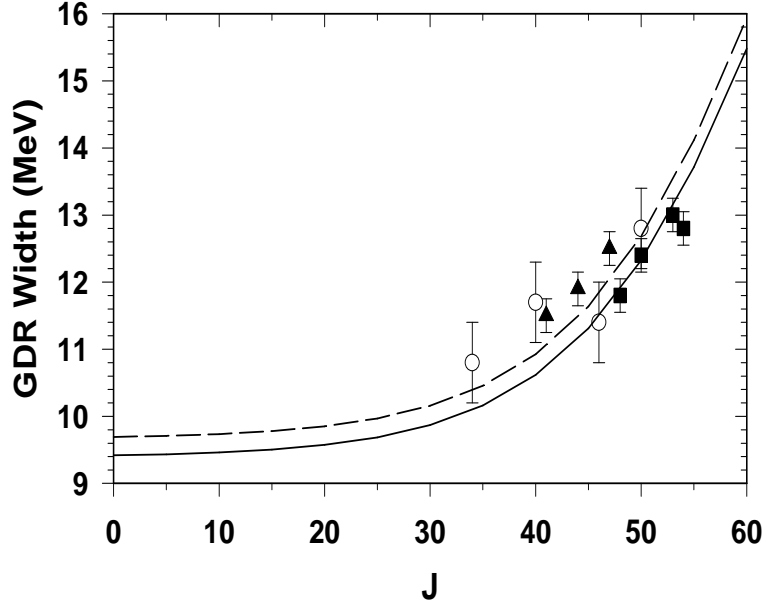}%
\caption{\label{figj} GDR widths are plotted against J and 
compared 
with Kusnezov calculation for T = 1.85 MeV (solid line) and 2.0 MeV 
(dashed line). The filled squares and triangles are data from this 
work while the points with the open circles are from experiment done by 
Bracco et al.~\cite{Bracco} (Details are described in the text).}
\end{figure}

\begin{figure}
\vspace{4 cm}
\includegraphics[height=8 cm, width=10 cm]{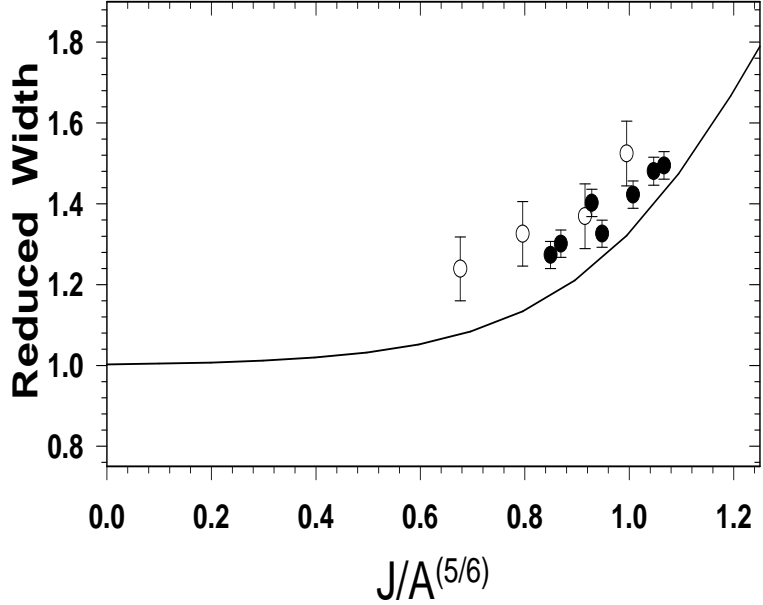}%
\caption{\label{figr} Reduced GDR widths are plotted against 
$\xi = J/A^{5/6}$ and compared with Kusnezov calculation. The 
filled circles are data from this work while the points with the 
open circles are from experiment done by Bracco et al.~\cite{Bracco} }
\end{figure}

There exist an earlier measurement of the GDR width 
from $^{109,110}$Sn at similar temperature and angular
 momentum by Bracco et al.~\cite{Bracco}. A reasonable 
agreement between those sets of data and predictions by 
Kusnezov parametrization has been shown in~\cite{Thon}. 
But 
recently by employing a new scheme of analysis a large 
mismatch
 between the data and the prediction has been seen~\cite{DRC}.  
The average values of T and J have been estimated for 
this data 
set also following the approach adopted in this work and 
is 
compared with the corresponding TSFM predictions. The 
results 
are shown in Figs. \ref{figt}, \ref{figj} \& \ref{figr} as open
 circles (see Table \ref{tab:brac1} for details of the 
parameters). 
While the temperature dependence is well-described by the 
TSFM 
calculation, the dependence of GDR width on J is 
under-predicted as 
shown in the Fig. \ref{figj} (for some points it is more 
than the 
error bar).
 
In Tables \ref{tab:kus} and \ref{tab:brac1} the errors in T 
indicate the average
temperature ranges associated to 80$\%$ (lower value) and 
20$\%$ (upper value) of the
total high energy $\gamma$-yield in CN decay chain.

The averaging scheme followed in this work, considers only the 
high energy $\gamma$-rays
which really influence the GDR width in the CN decay chain and 
neglects the remaining 
$\gamma$-decay cascade. In this low E$^*$ region, the average 
values of T do not differ
too much with the initial CN values primarily due to the emission 
of GDR $\gamma$-rays in
the first few steps of the CN decay cascade. As GDR decay is an 
average process, the averaging
of E$^*$, T, A and J is important in the CN decay chain. The change
of the value of A after averaging from that of the CN is not 
expected to make a significant
change in the conclusion. But the same is not true for J. Even a 
change in J by a few units,
by the averaging, can indeed alter the reduced width measurements.

The GDR widths measured in this work are also plotted 
against 
the average values of J's along with the theoretical predictions 
for 
T = 1.85 MeV and 2.0 MeV (solid and dashed lines respectively in the 
Fig. \ref{figj}). 
The data points represented by solid squares and triangles are 
for T $\leq$ 
1.85 MeV and 1.85 $<$ T $<$ 2.0 MeV respectively. They are agreeing 
well 
with Kusnezov parametrization. If the CN decay widths are considered 
(not shown in the figure), the matching improves marginally.

\begingroup
\squeezetable
\begin{table}
\caption{\label{tab:brac1}Table showing recalculation of parameters 
in this paper for different beam energies calculated by CASCADE for
 the experiment performed by Bracco et al.(see~\cite{Bracco})}
\begin{ruledtabular}
\begin{tabular}{|c|c|c|c|c|c|c|}
\vline
  E$_{beam}$ & J$_{CN}$ & J$_{mean}$ & T$_{mean}$ & E$_{GDR}$ & $\Gamma_{GDR}$ & FWHM in J \\ 
  (MeV)      & ($\hbar$) &  ($\hbar$) &   (MeV)   & (MeV)     &   (MeV)        & ($\hbar$)\\ \hline

223 & 44.0 & 40.0   & 1.76\/$^{+0.15}_{-0.02}$ & 15.0 $\pm$ 0.5 & 11.7 $\pm$ 0.6 & 16  \\
223 & 54.0 & 50.0   & 1.57$^{+0.12}_{-0.01}$ & 14.7 $\pm$ 0.5 & 12.8 $\pm$ 0.6 & 14\\
203 & 40.0 & 34.0 & 1.60$^{+0.13}_{-0.05}$  & 15.7 $\pm$ 0.5 & 10.8 $\pm$ 0.6 & 18\\
203 & 49.0 & 46.0 & 1.40$^{+0.08}_{-0.03}$  & 15.6 $\pm$ 0.5 & 11.4 $\pm$ 0.6 & 16 \\
\end{tabular}
\end{ruledtabular}
\end{table}
\endgroup

In the Fig. \ref{figr}, reduced GDR widths are plotted for this 
experiment as well as for the other set of existing data. The solid 
line corresponds to the global phenomenological free energy surface 
calculation shown in ~\cite{Kus} at T = 1.8 MeV within the TSFM. 
Reduced GDR widths $\Gamma_{red}$ for this experiment are shown in 
the figure by filled circles and the data for Bracco et al. are 
shown by open circles. Those reduced GDR widths are calculated 
using the parametrization, 
\begin{displaymath}
\Gamma_{red} = \Gamma_{exp}(T,J,A)/\Gamma_{theory}(T,J=0,A)^{(T+3T_0)/(4T_0)}
\end{displaymath}
and plotted against $\xi = \overline{J}/\overline{A}^{5/6}$, 
where T$_0$ = 1 MeV and
ground state GDR width $\Gamma_0$ = 3.8 MeV. Although the simple 
parametrization proposed
by Kusnezov reproduces reasonably well the trend of the experimental
 findings as a
function of spin, it would be interesting to see whether the more 
detailed calculation
with the Thermal Shape Fluctuation Model for specific nuclei can 
improve further the agreement with the data in reduced width plot for higher spins.

\section{Summary and Conclusion}
The GDR width built on excited states of $^{113}$Sb has been studied 
in the interval J = 40-60$\hbar$ and at a temperature ranging from 
T = 1.7-2.0 MeV using a part of the LAMBDA array. To decouple the 
effect of T and J from each other on GDR width, multiplicity detector 
assembly has been used. The angular momentum information was obtained 
from the low energy $\gamma$-multiplicity filter. Pre-equilibrium 
corrections on the excitation energies were performed. The temperature 
has been found from the average values of corrected excitation energy 
and angular momentum which provided a more stringent test for the 
existing TSFM for explaining the systematics of GDR width. Another set 
of existing data from earlier measurement of GDR width of Sn nuclei 
have been used for comparison.  In spite of the fact that the simple 
parametrization of Kusnezov describes the trend well, there remains a 
scope of detail TSFM calculation for further improvement in the 
parametrization of GDR width with J. Keeping J constant, the increase 
of $\Gamma$ with T, for T $\leq$ 2.0 MeV, is well understood by TSFM 
although the matching improves to some extent by the addition of CN 
decay width. 
So for a complete understanding of thermal shape fluctuation and 
parametrization of GDR width, more experiments in this mass range 
are needed. Also in the higher temperature region it is important to 
find out the detail pre-equilibrium energy loss involved.


\begin{acknowledgments}
The authors wish to thank I.Dioszegi and D.Hoffman for their help and suggestion
regarding the modification of CASCADE.
\end{acknowledgments}



%


\end{document}